\documentclass[11pt,a4paper]{article}
\pdfoutput=1

\usepackage{jheppub}
\usepackage{amsmath}
\usepackage{tabularx}
\usepackage{cancel}
\usepackage{changepage}
\usepackage[dvipsnames]{xcolor}

\newcommand{\al}{\alpha}
\newcommand{\be}{\beta}

\newcommand{\eps}{\epsilon}

\newcommand{\GeV}{\, \mathrm{GeV}}

\newcommand{\pt}{\partial}

\newcommand{\suc}{\ensuremath{ SU(3)_\text{c} }}
\newcommand{\sul}{\ensuremath{ SU(2)_\text{L} }}
\newcommand{\uy}{\ensuremath{ U(1)_\text{Y} }}
\newcommand{\loren}{\ensuremath{ SO(3,1) }}

\title{An operator basis for the Standard Model with an added scalar singlet}
\author[a]{Ben Gripaios}
\author[a,b]{and Dave Sutherland}

\affiliation[a]{Cavendish Laboratory, \\ J.J.\ Thomson Avenue, Cambridge, UK}
\affiliation[b]{Kavli Institute for Theoretical Physics, \\ UCSB Kohn Hall, Santa Barbara CA, USA}

\emailAdd{gripaios@hep.phy.cam.ac.uk}
\emailAdd{dws28@cam.ac.uk}

\preprint{Cavendish-HEP-16/05,NSF-KITP-16-056}

\abstract{Motivated by the possible di-gamma resonance at 750 GeV, we present a basis of effective operators for the Standard Model plus a scalar singlet at dimensions 5, 6, and 7. We point out that an earlier list at dimensions 5 and 6 contains two redundant operators at dimension 5.}


\arxivnumber{}

\begin{document} 
\maketitle
\section{Introduction \label{sec:int}}

Both ATLAS~\cite{atlasDiphoton} and CMS~\cite{CMS:2015dxe} have recently observed a slight excess in the $\gamma\gamma$ spectrum at an invariant mass of approximately $750 \GeV$. If this excess persists in the 2016 dataset, it will be apt to consider the relevant data in the most model independent way possible; any such approach will be invariably based on the number and nature of putative lagrangian terms which are consistent with the assumed symmetries of a new particle. Therefore, we present here a basis of operators up to and including mass dimension 7 for the simplest case: we assume Standard Model symmetries acting on the fields of the Standard Model plus a real scalar singlet, which we denote by $\phi$. We assume flavour diagonal couplings of the fermions. At dimensions 5 and 6, a putative basis has been given recently in \cite{Franceschini:2016gxv}. As we shall show, two linear combinations of the operators given there at dimension 5 are redundant (in the sense that they do not contribute to scattering amplitudes and can hence be eliminated from the basis). We also present a list of operators at dimension 7. As the large couplings required for the anomaly in $\gamma\gamma$ suggest that the cutoff scale of its effective field theory may be quite low, and as the number of basis operators at dimension 7 is still manageable, we reason that this list may also be useful.

In contrast to some recent elegant methods~\cite{Lehman:2015via,Henning:2015daa,Lehman:2015coa,Henning:2015alf} using Hilbert series, we proceed by brute force, using a computer to enumerate both the possible terms consistent with given symmetries, and the redundancies between them. Although this approach taxes the computer at higher dimensions due to the exponential growth of terms and relations between them, it is tractable at the dimensions of interest, and, moreover, the code is generic and yields a wealth of explicit information on the possible choices of basis, and how to convert to or inbetween arbitrary bases, for any particular input of fields and $SU(N)$-like symmetries. We will study these expressions systematically in a few interesting cases, including the Standard Model, in a forthcoming paper~\cite{deft}. However, for now, we just summarise our approach for the Standard Model plus singlet in Section~\ref{sec:method}, and then pick one basis according to simple criteria (such as minimising the number of derivatives in operators), writing it down in \S\ref{sec:results}. We compare with earlier results at dimensions 5 and 6 in \S\ref{sec:compare}.

\section{Method \label{sec:method}}

We use Python code to exhaustively generate all products of fields of a given mass dimension that transform as a singlet under the Standard Model gauge and Lorentz symmetries. Thereafter, the program enumerates all linear combinations of operators which do not contribute to scattering amplitudes at the same mass dimension order; finally, it performs some simple linear algebra to determine a suitable basis of operators to parametrise measurable effects in scattering amplitudes. We postpone a fuller exposition of the program to a forthcoming paper~\cite{deft}, and provide instead the following summary.

We consider the Standard Model fields, plus an additional scalar singlet, in irreps of the group $\suc \times \sul \times \uy \times \loren$. We parametrise the components of the irreps of the non-abelian $SU(N)$ gauge groups by upper indices (which transform in the fundamental irrep), lower indices (which transform in the conjugate irrep to the fundamental), and symmetric, anti-symmetric and traceless combinations thereof. Under Hermitian conjugation of the field, upper indices are lowered and \emph{vice versa}. We treat the Lorentz group irreps as if they were the irreps of the direct product of two $SU(2)$ groups, with the exception that, under Hermitian conjugation, their indices are swapped (\emph{i.e.}, undotted indices are dotted and \emph{vice versa}). As we are considering only flavour diagonal couplings, we work with just one fermion generation (flavour indices may be trivially reinserted). Table~\ref{tab:fieldcontent} displays the fields in component form.\footnote{In this formalism, the gauge boson field strengths are three component complex fields, \emph{e.g.} $B_{\al\be}$. We relate them to the more familiar $B_{\mu\nu}$ via the definition $\sigma^\mu_{\al \dot \al} \sigma^\nu_{\be \dot\be} B_{\mu \nu} = -\frac{1}{2} (\eps_{\dot\al \dot\be} B_{\al \be} + \eps_{\al \be} B^\dagger_{\dot\al \dot\be} )$, wherein all other quantities follow the conventions of~\cite{Dreiner:2008tw}.}

\begin{table}
\begin{tabular}{c | c | c | c | c | c | c |}
Field & Dimension & \suc & \sul & \uy & $SU(2)_\text{lor,L}$ & $SU(2)_\text{lor,R}$ \\ \hline \hline
${L_L}^{\al a}$ & $\frac{3}{2}$ & $1$ & $2$ & $-\frac{1}{2}$ & $2$ & $1$ \\
${e_R}^{\dot\al}$ & $\frac{3}{2}$ & $1$ & $1$ & $-1$ & $1$ & $2$ \\
${Q_L}^{\al a A}$ & $\frac{3}{2}$ & $3$ & $2$ & $\frac{1}{6}$ & $2$ & $1$ \\
${u_R}^{\dot\al A}$ & $\frac{3}{2}$ & $3$ & $1$ & $\frac{1}{3}$ & $1$ & $2$ \\
${d_R}^{\dot\al A}$ & $\frac{3}{2}$ & $3$ & $1$ & $-\frac{2}{3}$ & $1$ & $2$ \\
$H^{a}$ & $1$ & $1$ & $2$ & $\frac{1}{2}$ & $1$ & $1$ \\
$B_{(\al \be)}$ & $2$ & $1$ & $1$ & $0$ & $3$ & $1$ \\
$W^a_{b (\al \be)}$, $W^a_{a (\al \be)} \equiv 0$ & $2$ & $1$ & $3$ & $0$ & $3$ & $1$ \\
$G^A_{B (\al \be)}$, $G^A_{A (\al \be)} \equiv 0$ & $2$ & $8$ & $1$ & $0$ & $3$ & $1$ \\
$\phi$ & $1$ & $1$ & $1$ & $0$ & $1$ & $1$ \\
\end{tabular}
\caption{The fields in component form, plus their representations under the Standard Model symmetries. Key: $A,B,\ldots=SU(3)_c$, $a,b,\ldots=SU(2)_L$, $\alpha,\beta,\ldots=SU(2)_\text{lor,L}$, $\dot\alpha,\dot\beta,\ldots=SU(2)_\text{lor,R}$. \label{tab:fieldcontent}}
\end{table}

The only three non-Abelian group invariants are the Kronecker delta, which contracts an upper and lower index of the same group; the lower index Levi-Civita epsilon, which contracts $N$ $SU(N)$ upper indices; and, similarly, the upper index Levi-Civita epsilon. The program generates, up to a specified dimension, all products of fields and their covariant derivatives with zero net \uy~ charge. It then finds all possible ways of partitioning the upper and lower indices amongst epsilon and delta tensors. The resulting list of possible gauge and Lorentz singlet operators provide a hugely redundant description of measurable effects in scattering amplitudes. The program enumerates the following linear combinations of operators which make no such contribution to observables: Fierz relations (achieved by replacing products of epsilons with sums of products of deltas); integration by parts relations (by moving covariant derivatives amongst the fields of an operator); commuting derivatives (by switching the order of covariant derivatives and setting the difference equal to a sum of field strengths); and equation of motion relations, for operators of mass dimension greater than the spacetime dimension (by functionally differentiating the marginal operators and contracting the resulting equation of motion with all possible combinations of fields and derivatives).

For each linear combination the program constructs a row vector out of its coefficients, representing an unmeasurable direction in the vector space of operators. The position of each term's coefficient in the row vector corresponds to the ordering of the terms under an arbitrary comparison operator, such that undesirable terms that we hope to eliminate from the basis are to the left, and desirable terms that we wish to keep are to the right. Listed in descending priority, our preferences are for fewest derivatives, followed by fewest epsilons, followed by fewest group indices.\footnote{Up to simple IBP and Fierz rearrangements, this effectively reproduces the `Warsaw basis'~\cite{Grzadkowski:2010es} in the case of the Standard Model at dimension 6.}

The row vectors are combined into the rows of a matrix, whose rank indicates how many operators in our original list can be discarded to leave a suitable basis. The program then puts the matrix in reduced row echelon form. Each operator whose coefficient is the first non-zero entry in a row of the reduced matrix may then be consistently eliminated.

We have verified the results of the program against a few simple examples, including the model documented in Appendix B of~\cite{Einhorn:2013kja}, as well as the one generation Standard Model at dimensions 5~\cite{PhysRevLett.43.1566}, 6~\cite{Grzadkowski:2010es}, and 7~\cite{Lehman:2015coa}.

\section{Results \label{sec:results}}

Table~\ref{tab:operatorCount} shows the number of independent operators at each mass dimension up to 7 for a one generation Standard Model plus a real scalar singlet. Note that we count an operator and its distinct Hermitian conjugate separately, and the entries hence reflect the number of real parameters that can be fit at each dimension, not accounting for the freedom to canonically normalise fields, nor to perform subsequent unitary/orthogonal transformations thereof.

\begin{table}
\centering
\begin{tabular}{l c c c c c c c}
$d$ &1&2&3&4&5&6&7\\ \hline
SM $+$ singlet & 1  & 2 &  2 & 19 & 17 & 102 & 176 \\
~ of which SM & 0 &  1 &  0 & 16 &  2 & 84 & 30 \\
~ of which new & 1  & 1 &  2 &  3 & 15 & 18 & 146 \\
\end{tabular}
\caption{The number of independent operators at each dimension $d \leq 7$ that contribute to scattering amplitudes. In the last two rows we partition them according to whether they do not or do contain a $\phi$ field respectively.\label{tab:operatorCount}}
\end{table}

A suitable basis of operators consists of the usual Standard Model operators at $d=2$ and $d=4$, the Weinberg operator at $d=5$, the operators of Tables 2 and 3 of \cite{Grzadkowski:2010es}\footnote{excluding $Q^{(3)}_{qqq}$ and its Hermitian conjugate, which are not independent when there is only one matter generation} at $d=6$, and the operators of Appendix D of \cite{Lehman:2015coa}\footnote{The operator denoted $L^3 e_c H$ should have multiplicity 1 when $N_f = 1$.} at $d=7$, plus their distinct Hermitian conjugates, plus the following at each dimension $d \leq 7$ that include at least one scalar singlet $\phi$.

\begin{adjustwidth*}{3cm}{-1cm}
\begin{description}
\item[$d=1$] \textcolor{red}{ $\phi$ }
\item[$d=2$] $\phi \phi$ 
\item[$d=3$] \textcolor{red}{ $ {H^\dagger}^{}_{a } {H}^{a }_{} {\phi}^{}_{}$ }, \textcolor{red}{ 
$ {\phi}^{}_{} {\phi}^{}_{} {\phi}^{}_{}$ }
\item[$d=4$] $\epsilon^{\alpha \beta }_{} \epsilon^{\dot\alpha \dot\beta }_{} D_{\alpha \dot\beta} {\phi}^{}_{} D_{\beta \dot\alpha} {\phi}^{}_{}$, 
$ {H^\dagger}^{}_{a } {H}^{a }_{} {\phi}^{}_{} {\phi}^{}_{}$, 
$ {\phi}^{}_{} {\phi}^{}_{} {\phi}^{}_{} {\phi}^{}_{}$
\item[$d=5$]\textcolor{ForestGreen}{ $\epsilon^{\alpha \beta }_{} \epsilon^{\gamma \delta }_{} {W}^{a }_{\gamma \beta b } {W}^{b }_{\alpha \delta a } {\phi}^{}_{}$ }, 
\textcolor{ForestGreen}{ $\epsilon^{\dot\alpha \dot\beta }_{} \epsilon^{\dot\gamma \dot\delta }_{} {W^\dagger}^{a }_{\dot\delta \dot\beta b } {W^\dagger}^{b }_{\dot\alpha \dot\gamma a } {\phi}^{}_{}$ }, 
\textcolor{ForestGreen}{ $\epsilon^{\alpha \beta }_{} \epsilon^{\gamma \delta }_{} {G}^{A }_{\gamma \beta B } {G}^{B }_{\alpha \delta A } {\phi}^{}_{}$ }, 
\textcolor{ForestGreen}{ $\epsilon^{\dot\alpha \dot\beta }_{} \epsilon^{\dot\gamma \dot\delta }_{} {G^\dagger}^{A }_{\dot\delta \dot\beta B } {G^\dagger}^{B }_{\dot\alpha \dot\gamma A } {\phi}^{}_{}$ }, 
\textcolor{ForestGreen}{ $\epsilon^{}_{\dot\alpha \dot\beta } \epsilon^{a b }_{} {H^\dagger}^{}_{a } {Q_L^\dagger}^{\dot\beta }_{A b } {u_R}^{A \dot\alpha }_{} {\phi}^{}_{}$ }, 
\textcolor{ForestGreen}{ $\epsilon^{}_{\alpha \beta } \epsilon^{}_{a b } {H}^{a }_{} {Q_L}^{\alpha A b }_{} {u_R^\dagger}^{\beta }_{A } {\phi}^{}_{}$ }, 
\textcolor{ForestGreen}{ $\epsilon^{\alpha \beta }_{} \epsilon^{\gamma \delta }_{} {B}^{}_{\gamma \beta } {B}^{}_{\alpha \delta } {\phi}^{}_{}$ }, 
\textcolor{ForestGreen}{ $\epsilon^{\dot\alpha \dot\beta }_{} \epsilon^{\dot\gamma \dot\delta }_{} {B^\dagger}^{}_{\dot\delta \dot\beta } {B^\dagger}^{}_{\dot\alpha \dot\gamma } {\phi}^{}_{}$ }, 
\textcolor{ForestGreen}{ $\epsilon^{}_{\dot\alpha \dot\beta } {H}^{a }_{} {Q_L^\dagger}^{\dot\beta }_{A a } {d_R}^{A \dot\alpha }_{} {\phi}^{}_{}$ }, 
\textcolor{ForestGreen}{ $\epsilon^{}_{\alpha \beta } {H^\dagger}^{}_{a } {Q_L}^{\alpha A a }_{} {d_R^\dagger}^{\beta }_{A } {\phi}^{}_{}$ }, 
\textcolor{ForestGreen}{ $\epsilon^{}_{\alpha \beta } {H^\dagger}^{}_{a } {e_R^\dagger}^{\alpha }_{} {L_L}^{\beta a }_{} {\phi}^{}_{}$ }, 
\textcolor{ForestGreen}{ $\epsilon^{}_{\dot\alpha \dot\beta } {H}^{a }_{} {e_R}^{\dot\beta }_{} {L_L^\dagger}^{\dot\alpha }_{a } {\phi}^{}_{}$ }, 
\textcolor{red}{ $ {H^\dagger}^{}_{a } {H^\dagger}^{}_{b } {H}^{a }_{} {H}^{b }_{} {\phi}^{}_{}$ }, 
\textcolor{red}{ $ {H^\dagger}^{}_{a } {H}^{a }_{} {\phi}^{}_{} {\phi}^{}_{} {\phi}^{}_{}$ }, 
\textcolor{red}{ $ {\phi}^{}_{} {\phi}^{}_{} {\phi}^{}_{} {\phi}^{}_{} {\phi}^{}_{}$ }

\item[$d=6$] \textcolor{blue}{ $\epsilon^{\dot\alpha \dot\beta }_{} \epsilon^{\dot\gamma \dot\delta }_{} {B^\dagger}^{}_{\dot\delta \dot\beta } {B^\dagger}^{}_{\dot\alpha \dot\gamma } {\phi}^{}_{} {\phi}^{}_{}$ }, 
\textcolor{blue}{ $\epsilon^{\alpha \beta }_{} \epsilon^{\gamma \delta }_{} {B}^{}_{\gamma \beta } {B}^{}_{\alpha \delta } {\phi}^{}_{} {\phi}^{}_{}$ }, 
\textcolor{blue}{ $\epsilon^{\dot\alpha \dot\beta }_{} \epsilon^{\dot\gamma \dot\delta }_{} {G^\dagger}^{A }_{\dot\delta \dot\beta B } {G^\dagger}^{B }_{\dot\alpha \dot\gamma A } {\phi}^{}_{} {\phi}^{}_{}$ }, 
\textcolor{blue}{ $\epsilon^{\alpha \beta }_{} \epsilon^{\gamma \delta }_{} {G}^{A }_{\gamma \beta B } {G}^{B }_{\alpha \delta A } {\phi}^{}_{} {\phi}^{}_{}$ }, 
$ {H^\dagger}^{}_{a } {H^\dagger}^{}_{b } {H}^{a }_{} {H}^{b }_{} {\phi}^{}_{} {\phi}^{}_{}$, 
$ {H^\dagger}^{}_{a } {H}^{a }_{} {\phi}^{}_{} {\phi}^{}_{} {\phi}^{}_{} {\phi}^{}_{}$, 
\textcolor{blue}{ $\epsilon^{}_{\dot\alpha \dot\beta } {H}^{a }_{} {e_R}^{\dot\beta }_{} {L_L^\dagger}^{\dot\alpha }_{a } {\phi}^{}_{} {\phi}^{}_{}$ }, 
\textcolor{blue}{ $\epsilon^{}_{\alpha \beta } {H^\dagger}^{}_{a } {e_R^\dagger}^{\alpha }_{} {L_L}^{\beta a }_{} {\phi}^{}_{} {\phi}^{}_{}$ }, 
\textcolor{blue}{ $\epsilon^{}_{\alpha \beta } {H^\dagger}^{}_{a } {Q_L}^{\alpha A a }_{} {d_R^\dagger}^{\beta }_{A } {\phi}^{}_{} {\phi}^{}_{}$ }, 
\textcolor{blue}{ $\epsilon^{}_{\dot\alpha \dot\beta } {H}^{a }_{} {Q_L^\dagger}^{\dot\beta }_{A a } {d_R}^{A \dot\alpha }_{} {\phi}^{}_{} {\phi}^{}_{}$ }, 
\textcolor{blue}{ $\epsilon^{}_{\dot\alpha \dot\beta } \epsilon^{a b }_{} {H^\dagger}^{}_{a } {Q_L^\dagger}^{\dot\beta }_{A b } {u_R}^{A \dot\alpha }_{} {\phi}^{}_{} {\phi}^{}_{}$ }, 
\textcolor{blue}{ $\epsilon^{}_{\alpha \beta } \epsilon^{}_{a b } {H}^{a }_{} {Q_L}^{\alpha A b }_{} {u_R^\dagger}^{\beta }_{A } {\phi}^{}_{} {\phi}^{}_{}$ }, 
$ {\phi}^{}_{} {\phi}^{}_{} {\phi}^{}_{} {\phi}^{}_{} {\phi}^{}_{} {\phi}^{}_{}$, 
\textcolor{blue}{ $\epsilon^{\alpha \beta }_{} \epsilon^{\gamma \delta }_{} {W}^{a }_{\gamma \beta b } {W}^{b }_{\alpha \delta a } {\phi}^{}_{} {\phi}^{}_{}$ }, 
\textcolor{blue}{ $\epsilon^{\dot\alpha \dot\beta }_{} \epsilon^{\dot\gamma \dot\delta }_{} {W^\dagger}^{a }_{\dot\delta \dot\beta b } {W^\dagger}^{b }_{\dot\alpha \dot\gamma a } {\phi}^{}_{} {\phi}^{}_{}$ }, 
$\epsilon^{\alpha \beta }_{} \epsilon^{\dot\alpha \dot\beta }_{} {H^\dagger}^{}_{a } {H}^{a }_{} D_{\alpha \dot\beta} {\phi}^{}_{} D_{\beta \dot\alpha} {\phi}^{}_{}$

\line(1,0){40}

\textcolor{ForestGreen}{ $\epsilon^{}_{\dot\alpha \dot\beta } \epsilon^{a b }_{} \epsilon^{d e }_{} {H^\dagger}^{}_{d } {H^\dagger}^{}_{b } {L_L^\dagger}^{\dot\beta }_{a } {L_L^\dagger}^{\dot\alpha }_{e } {\phi}^{}_{}$ }, 
\textcolor{ForestGreen}{ $\epsilon^{}_{\alpha \beta } \epsilon^{}_{a b } \epsilon^{}_{d e } {H}^{d }_{} {H}^{b }_{} {L_L}^{\alpha a }_{} {L_L}^{\beta e }_{} {\phi}^{}_{}$ }

\item[$d=7$]

\textcolor{ForestGreen}{ $\epsilon^{\alpha \beta }_{} \epsilon^{\gamma \delta }_{} {B}^{}_{\gamma \beta } {B}^{}_{\alpha \delta } {\phi}^{}_{} {\phi}^{}_{} {\phi}^{}_{}$ }, 
\textcolor{ForestGreen}{ $\epsilon^{\dot\alpha \dot\beta }_{} \epsilon^{\dot\gamma \dot\delta }_{} {B^\dagger}^{}_{\dot\delta \dot\beta } {B^\dagger}^{}_{\dot\alpha \dot\gamma } {\phi}^{}_{} {\phi}^{}_{} {\phi}^{}_{}$ }, 
\textcolor{ForestGreen}{ $\epsilon^{\dot\alpha \dot\beta }_{} \epsilon^{\dot\gamma \dot\delta }_{} \epsilon^{\dot\mu \dot\nu }_{} {G^\dagger}^{A }_{\dot\nu \dot\delta C } {G^\dagger}^{B }_{\dot\beta \dot\gamma A } {G^\dagger}^{C }_{\dot\mu \dot\alpha B } {\phi}^{}_{}$ }, 
\textcolor{ForestGreen}{ $\epsilon^{\alpha \beta }_{} \epsilon^{\gamma \delta }_{} \epsilon^{\mu \nu }_{} {G}^{A }_{\mu \beta C } {G}^{B }_{\delta \nu A } {G}^{C }_{\alpha \gamma B } {\phi}^{}_{}$ }, 
\textcolor{ForestGreen}{ $\epsilon^{\alpha \beta }_{} \epsilon^{\gamma \delta }_{} {G}^{A }_{\gamma \beta B } {G}^{B }_{\alpha \delta A } {\phi}^{}_{} {\phi}^{}_{} {\phi}^{}_{}$ }, 
\textcolor{ForestGreen}{ $\epsilon^{\dot\alpha \dot\beta }_{} \epsilon^{\dot\gamma \dot\delta }_{} {G^\dagger}^{A }_{\dot\delta \dot\beta B } {G^\dagger}^{B }_{\dot\alpha \dot\gamma A } {\phi}^{}_{} {\phi}^{}_{} {\phi}^{}_{}$ }, 
\textcolor{ForestGreen}{ $ {H^\dagger}^{}_{a } {B}^{}_{\alpha \beta } {e_R^\dagger}^{\alpha }_{} {L_L}^{\beta a }_{} {\phi}^{}_{}$ }, 
\textcolor{ForestGreen}{ $ {H}^{a }_{} {B^\dagger}^{}_{\dot\alpha \dot\beta } {e_R}^{\dot\alpha }_{} {L_L^\dagger}^{\dot\beta }_{a } {\phi}^{}_{}$ }, 
\textcolor{ForestGreen}{ $ {H}^{a }_{} {B^\dagger}^{}_{\dot\alpha \dot\beta } {Q_L^\dagger}^{\dot\alpha }_{A a } {d_R}^{A \dot\beta }_{} {\phi}^{}_{}$ }, 
\textcolor{ForestGreen}{ $ {H^\dagger}^{}_{a } {B}^{}_{\alpha \beta } {Q_L}^{\alpha A a }_{} {d_R^\dagger}^{\beta }_{A } {\phi}^{}_{}$ }, 
\textcolor{ForestGreen}{ $\epsilon^{a b }_{} {H^\dagger}^{}_{a } {B^\dagger}^{}_{\dot\alpha \dot\beta } {Q_L^\dagger}^{\dot\alpha }_{A b } {u_R}^{A \dot\beta }_{} {\phi}^{}_{}$ }, 
\textcolor{ForestGreen}{ $\epsilon^{}_{a b } {H}^{a }_{} {B}^{}_{\alpha \beta } {Q_L}^{\alpha A b }_{} {u_R^\dagger}^{\beta }_{A } {\phi}^{}_{}$ }, 
\textcolor{ForestGreen}{ $\epsilon^{\alpha \beta }_{} \epsilon^{\gamma \delta }_{} {H^\dagger}^{}_{a } {H}^{a }_{} {B}^{}_{\gamma \beta } {B}^{}_{\alpha \delta } {\phi}^{}_{}$ }, 
\textcolor{ForestGreen}{ $\epsilon^{\dot\alpha \dot\beta }_{} \epsilon^{\dot\gamma \dot\delta }_{} {H^\dagger}^{}_{a } {H}^{a }_{} {B^\dagger}^{}_{\dot\delta \dot\beta } {B^\dagger}^{}_{\dot\alpha \dot\gamma } {\phi}^{}_{}$ }, 
\textcolor{ForestGreen}{ $\epsilon^{\alpha \beta }_{} \epsilon^{\gamma \delta }_{} {H^\dagger}^{}_{b } {H}^{a }_{} {B}^{}_{\gamma \beta } {W}^{b }_{\alpha \delta a } {\phi}^{}_{}$ }, 
\textcolor{ForestGreen}{ $\epsilon^{\dot\alpha \dot\beta }_{} \epsilon^{\dot\gamma \dot\delta }_{} {H^\dagger}^{}_{b } {H}^{a }_{} {B^\dagger}^{}_{\dot\delta \dot\beta } {W^\dagger}^{b }_{\dot\alpha \dot\gamma a } {\phi}^{}_{}$ }, 
\textcolor{ForestGreen}{ $\epsilon^{\alpha \beta }_{} \epsilon^{\gamma \delta }_{} {H^\dagger}^{}_{a } {H}^{a }_{} {G}^{A }_{\gamma \beta B } {G}^{B }_{\alpha \delta A } {\phi}^{}_{}$ }, 
\textcolor{ForestGreen}{ $\epsilon^{\dot\alpha \dot\beta }_{} \epsilon^{\dot\gamma \dot\delta }_{} {H^\dagger}^{}_{a } {H}^{a }_{} {G^\dagger}^{A }_{\dot\delta \dot\beta B } {G^\dagger}^{B }_{\dot\alpha \dot\gamma A } {\phi}^{}_{}$ }, 
\textcolor{red}{ $ {H^\dagger}^{}_{a } {H^\dagger}^{}_{b } {H^\dagger}^{}_{d } {H}^{a }_{} {H}^{b }_{} {H}^{d }_{} {\phi}^{}_{}$ }, 
\textcolor{red}{ $ {H^\dagger}^{}_{a } {H^\dagger}^{}_{b } {H}^{a }_{} {H}^{b }_{} {\phi}^{}_{} {\phi}^{}_{} {\phi}^{}_{}$ }, 
\textcolor{ForestGreen}{ $\epsilon^{}_{\alpha \beta } {H^\dagger}^{}_{a } {H^\dagger}^{}_{b } {H}^{a }_{} {e_R^\dagger}^{\alpha }_{} {L_L}^{\beta b }_{} {\phi}^{}_{}$ }, 
\textcolor{ForestGreen}{ $\epsilon^{}_{\dot\alpha \dot\beta } {H^\dagger}^{}_{a } {H}^{a }_{} {H}^{b }_{} {e_R}^{\dot\beta }_{} {L_L^\dagger}^{\dot\alpha }_{b } {\phi}^{}_{}$ }, 
\textcolor{ForestGreen}{ $\epsilon^{}_{\dot\alpha \dot\beta } {H^\dagger}^{}_{a } {H}^{a }_{} {H}^{b }_{} {Q_L^\dagger}^{\dot\beta }_{A b } {d_R}^{A \dot\alpha }_{} {\phi}^{}_{}$ }, 
\textcolor{ForestGreen}{ $\epsilon^{}_{\alpha \beta } {H^\dagger}^{}_{a } {H^\dagger}^{}_{b } {H}^{a }_{} {Q_L}^{\alpha A b }_{} {d_R^\dagger}^{\beta }_{A } {\phi}^{}_{}$ }, 
\textcolor{ForestGreen}{ $\epsilon^{}_{\dot\alpha \dot\beta } \epsilon^{a b }_{} {H^\dagger}^{}_{d } {H^\dagger}^{}_{b } {H}^{d }_{} {Q_L^\dagger}^{\dot\beta }_{A a } {u_R}^{A \dot\alpha }_{} {\phi}^{}_{}$ }, 
\textcolor{ForestGreen}{ $\epsilon^{}_{\alpha \beta } \epsilon^{}_{a b } {H^\dagger}^{}_{d } {H}^{d }_{} {H}^{b }_{} {Q_L}^{\alpha A a }_{} {u_R^\dagger}^{\beta }_{A } {\phi}^{}_{}$ }, 
\textcolor{red}{ $ {H^\dagger}^{}_{a } {H}^{a }_{} {\phi}^{}_{} {\phi}^{}_{} {\phi}^{}_{} {\phi}^{}_{} {\phi}^{}_{}$ }, 
\textcolor{ForestGreen}{ $\epsilon^{\alpha \beta }_{} \epsilon^{\gamma \delta }_{} {H^\dagger}^{}_{d } {H}^{a }_{} {W}^{b }_{\gamma \beta a } {W}^{d }_{\alpha \delta b } {\phi}^{}_{}$ }, 
\textcolor{ForestGreen}{ $\epsilon^{\dot\alpha \dot\beta }_{} \epsilon^{\dot\gamma \dot\delta }_{} {H^\dagger}^{}_{d } {H}^{a }_{} {W^\dagger}^{b }_{\dot\delta \dot\beta a } {W^\dagger}^{d }_{\dot\alpha \dot\gamma b } {\phi}^{}_{}$ }, 
\textcolor{ForestGreen}{ $\epsilon^{}_{\alpha \beta } {H^\dagger}^{}_{a } {e_R^\dagger}^{\alpha }_{} {L_L}^{\beta a }_{} {\phi}^{}_{} {\phi}^{}_{} {\phi}^{}_{}$ }, 
\textcolor{ForestGreen}{ $\epsilon^{}_{\dot\alpha \dot\beta } {H}^{a }_{} {e_R}^{\dot\beta }_{} {L_L^\dagger}^{\dot\alpha }_{a } {\phi}^{}_{} {\phi}^{}_{} {\phi}^{}_{}$ }, 
\textcolor{ForestGreen}{ $ {H^\dagger}^{}_{b } {e_R^\dagger}^{\alpha }_{} {L_L}^{\beta a }_{} {W}^{b }_{\alpha \beta a } {\phi}^{}_{}$ }, 
\textcolor{ForestGreen}{ $ {H}^{a }_{} {e_R}^{\dot\alpha }_{} {L_L^\dagger}^{\dot\beta }_{b } {W^\dagger}^{b }_{\dot\alpha \dot\beta a } {\phi}^{}_{}$ }, 
\textcolor{ForestGreen}{ $ {H^\dagger}^{}_{a } {Q_L}^{\alpha A a }_{} {d_R^\dagger}^{\beta }_{B } {G}^{B }_{\alpha \beta A } {\phi}^{}_{}$ }, 
\textcolor{ForestGreen}{ $ {H}^{a }_{} {Q_L^\dagger}^{\dot\alpha }_{B a } {d_R}^{A \dot\beta }_{} {G^\dagger}^{B }_{\dot\alpha \dot\beta A } {\phi}^{}_{}$ }, 
\textcolor{ForestGreen}{ $\epsilon^{}_{\dot\alpha \dot\beta } {H}^{a }_{} {Q_L^\dagger}^{\dot\beta }_{A a } {d_R}^{A \dot\alpha }_{} {\phi}^{}_{} {\phi}^{}_{} {\phi}^{}_{}$ }, 
\textcolor{ForestGreen}{ $\epsilon^{}_{\alpha \beta } {H^\dagger}^{}_{a } {Q_L}^{\alpha A a }_{} {d_R^\dagger}^{\beta }_{A } {\phi}^{}_{} {\phi}^{}_{} {\phi}^{}_{}$ }, 
\textcolor{ForestGreen}{ $ {H^\dagger}^{}_{b } {Q_L}^{\alpha A a }_{} {d_R^\dagger}^{\beta }_{A } {W}^{b }_{\alpha \beta a } {\phi}^{}_{}$ }, 
\textcolor{ForestGreen}{ $ {H}^{a }_{} {Q_L^\dagger}^{\dot\alpha }_{A b } {d_R}^{A \dot\beta }_{} {W^\dagger}^{b }_{\dot\alpha \dot\beta a } {\phi}^{}_{}$ }, 
\textcolor{ForestGreen}{ $\epsilon^{}_{a b } {H}^{a }_{} {Q_L}^{\alpha A b }_{} {u_R^\dagger}^{\beta }_{B } {G}^{B }_{\alpha \beta A } {\phi}^{}_{}$ }, 
\textcolor{ForestGreen}{ $\epsilon^{a b }_{} {H^\dagger}^{}_{a } {Q_L^\dagger}^{\dot\alpha }_{B b } {u_R}^{A \dot\beta }_{} {G^\dagger}^{B }_{\dot\alpha \dot\beta A } {\phi}^{}_{}$ }, 
\textcolor{ForestGreen}{ $\epsilon^{}_{\dot\alpha \dot\beta } \epsilon^{a b }_{} {H^\dagger}^{}_{a } {Q_L^\dagger}^{\dot\beta }_{A b } {u_R}^{A \dot\alpha }_{} {\phi}^{}_{} {\phi}^{}_{} {\phi}^{}_{}$ }, 
\textcolor{ForestGreen}{ $\epsilon^{}_{\alpha \beta } \epsilon^{}_{a b } {H}^{a }_{} {Q_L}^{\alpha A b }_{} {u_R^\dagger}^{\beta }_{A } {\phi}^{}_{} {\phi}^{}_{} {\phi}^{}_{}$ }, 
\textcolor{ForestGreen}{ $\epsilon^{}_{a b } {H}^{a }_{} {Q_L}^{\alpha A d }_{} {u_R^\dagger}^{\beta }_{A } {W}^{b }_{\alpha \beta d } {\phi}^{}_{}$ }, 
\textcolor{ForestGreen}{ $\epsilon^{a b }_{} {H^\dagger}^{}_{a } {Q_L^\dagger}^{\dot\alpha }_{A d } {u_R}^{A \dot\beta }_{} {W^\dagger}^{d }_{\dot\alpha \dot\beta b } {\phi}^{}_{}$ }, 
\textcolor{red}{ $ {\phi}^{}_{} {\phi}^{}_{} {\phi}^{}_{} {\phi}^{}_{} {\phi}^{}_{} {\phi}^{}_{} {\phi}^{}_{}$ }, 
\textcolor{ForestGreen}{ $\epsilon^{\alpha \beta }_{} \epsilon^{\gamma \delta }_{} {W}^{a }_{\gamma \beta b } {W}^{b }_{\alpha \delta a } {\phi}^{}_{} {\phi}^{}_{} {\phi}^{}_{}$ }, 
\textcolor{ForestGreen}{ $\epsilon^{\dot\alpha \dot\beta }_{} \epsilon^{\dot\gamma \dot\delta }_{} {W^\dagger}^{a }_{\dot\delta \dot\beta b } {W^\dagger}^{b }_{\dot\alpha \dot\gamma a } {\phi}^{}_{} {\phi}^{}_{} {\phi}^{}_{}$ }, 
\textcolor{ForestGreen}{ $\epsilon^{\dot\alpha \dot\beta }_{} \epsilon^{\dot\gamma \dot\delta }_{} \epsilon^{\dot\mu \dot\nu }_{} {W^\dagger}^{a }_{\dot\nu \dot\delta d } {W^\dagger}^{b }_{\dot\beta \dot\gamma a } {W^\dagger}^{d }_{\dot\mu \dot\alpha b } {\phi}^{}_{}$ }, 
\textcolor{ForestGreen}{ $\epsilon^{\alpha \beta }_{} \epsilon^{\gamma \delta }_{} \epsilon^{\mu \nu }_{} {W}^{a }_{\mu \beta d } {W}^{b }_{\delta \nu a } {W}^{d }_{\alpha \gamma b } {\phi}^{}_{}$ }, 
\textcolor{red}{ $\epsilon^{}_{\alpha \beta } \epsilon^{}_{\dot\alpha \dot\beta } {d_R^\dagger}^{\alpha }_{A } {d_R^\dagger}^{\beta }_{B } {d_R}^{A \dot\beta }_{} {d_R}^{B \dot\alpha }_{} {\phi}^{}_{}$ }, 
\textcolor{red}{ $\epsilon^{}_{\alpha \beta } \epsilon^{}_{\dot\alpha \dot\beta } {e_R^\dagger}^{\alpha }_{} {e_R}^{\dot\beta }_{} {d_R^\dagger}^{\beta }_{A } {d_R}^{A \dot\alpha }_{} {\phi}^{}_{}$ }, 
\textcolor{red}{ $\epsilon^{}_{\alpha \beta } \epsilon^{}_{\dot\alpha \dot\beta } {e_R^\dagger}^{\alpha }_{} {e_R^\dagger}^{\beta }_{} {e_R}^{\dot\beta }_{} {e_R}^{\dot\alpha }_{} {\phi}^{}_{}$ }, 
\textcolor{red}{ $\epsilon^{}_{\alpha \beta } \epsilon^{}_{\dot\alpha \dot\beta } {e_R^\dagger}^{\alpha }_{} {e_R}^{\dot\beta }_{} {L_L^\dagger}^{\dot\alpha }_{a } {L_L}^{\beta a }_{} {\phi}^{}_{}$ }, 
\textcolor{red}{ $\epsilon^{}_{\alpha \beta } \epsilon^{}_{\dot\alpha \dot\beta } {e_R^\dagger}^{\alpha }_{} {e_R}^{\dot\beta }_{} {Q_L^\dagger}^{\dot\alpha }_{A a } {Q_L}^{\beta A a }_{} {\phi}^{}_{}$ }, 
\textcolor{red}{ $\epsilon^{}_{\alpha \beta } \epsilon^{}_{\dot\alpha \dot\beta } {e_R^\dagger}^{\alpha }_{} {e_R}^{\dot\beta }_{} {u_R^\dagger}^{\beta }_{A } {u_R}^{A \dot\alpha }_{} {\phi}^{}_{}$ }, 
\textcolor{ForestGreen}{ $\epsilon^{}_{\alpha \beta } \epsilon^{}_{\dot\alpha \dot\beta } {e_R^\dagger}^{\alpha }_{} {L_L}^{\beta a }_{} {Q_L^\dagger}^{\dot\beta }_{A a } {d_R}^{A \dot\alpha }_{} {\phi}^{}_{}$ }, 
\textcolor{ForestGreen}{ $\epsilon^{}_{\alpha \beta } \epsilon^{}_{\dot\alpha \dot\beta } {e_R}^{\dot\beta }_{} {L_L^\dagger}^{\dot\alpha }_{a } {Q_L}^{\alpha A a }_{} {d_R^\dagger}^{\beta }_{A } {\phi}^{}_{}$ }, 
\textcolor{ForestGreen}{ $\epsilon^{}_{\dot\alpha \dot\beta } \epsilon^{}_{\dot\gamma \dot\delta } \epsilon^{a b }_{} {e_R}^{\dot\delta }_{} {L_L^\dagger}^{\dot\beta }_{a } {Q_L^\dagger}^{\dot\gamma }_{A b } {u_R}^{A \dot\alpha }_{} {\phi}^{}_{}$ }, 
\textcolor{ForestGreen}{ $\epsilon^{}_{\dot\alpha \dot\beta } \epsilon^{}_{\dot\gamma \dot\delta } \epsilon^{a b }_{} {e_R}^{\dot\delta }_{} {L_L^\dagger}^{\dot\gamma }_{a } {Q_L^\dagger}^{\dot\beta }_{A b } {u_R}^{A \dot\alpha }_{} {\phi}^{}_{}$ }, 
\textcolor{ForestGreen}{ $\epsilon^{}_{\alpha \beta } \epsilon^{}_{\gamma \delta } \epsilon^{}_{a b } {e_R^\dagger}^{\alpha }_{} {L_L}^{\delta a }_{} {Q_L}^{\beta A b }_{} {u_R^\dagger}^{\gamma }_{A } {\phi}^{}_{}$ }, 
\textcolor{ForestGreen}{ $\epsilon^{}_{\alpha \beta } \epsilon^{}_{\gamma \delta } \epsilon^{}_{a b } {e_R^\dagger}^{\alpha }_{} {L_L}^{\beta a }_{} {Q_L}^{\delta A b }_{} {u_R^\dagger}^{\gamma }_{A } {\phi}^{}_{}$ }, 
\textcolor{red}{ $\epsilon^{}_{\alpha \beta } \epsilon^{}_{\dot\alpha \dot\beta } {L_L^\dagger}^{\dot\beta }_{a } {L_L}^{\alpha a }_{} {d_R^\dagger}^{\beta }_{A } {d_R}^{A \dot\alpha }_{} {\phi}^{}_{}$ }, 
\textcolor{red}{ $\epsilon^{}_{\alpha \beta } \epsilon^{}_{\dot\alpha \dot\beta } {L_L^\dagger}^{\dot\beta }_{a } {L_L^\dagger}^{\dot\alpha }_{b } {L_L}^{\alpha a }_{} {L_L}^{\beta b }_{} {\phi}^{}_{}$ }, 
\textcolor{red}{ $\epsilon^{}_{\alpha \beta } \epsilon^{}_{\dot\alpha \dot\beta } {L_L^\dagger}^{\dot\beta }_{a } {L_L}^{\alpha a }_{} {Q_L^\dagger}^{\dot\alpha }_{A b } {Q_L}^{\beta A b }_{} {\phi}^{}_{}$ }, 
\textcolor{red}{ $\epsilon^{}_{\alpha \beta } \epsilon^{}_{\dot\alpha \dot\beta } {L_L^\dagger}^{\dot\beta }_{b } {L_L}^{\alpha a }_{} {Q_L^\dagger}^{\dot\alpha }_{A a } {Q_L}^{\beta A b }_{} {\phi}^{}_{}$ }, 
\textcolor{red}{ $\epsilon^{}_{\alpha \beta } \epsilon^{}_{\dot\alpha \dot\beta } {L_L^\dagger}^{\dot\beta }_{a } {L_L}^{\alpha a }_{} {u_R^\dagger}^{\beta }_{A } {u_R}^{A \dot\alpha }_{} {\phi}^{}_{}$ }, 
\textcolor{red}{ $\epsilon^{}_{\alpha \beta } \epsilon^{}_{\dot\alpha \dot\beta } {Q_L^\dagger}^{\dot\beta }_{A a } {Q_L}^{\alpha A a }_{} {d_R^\dagger}^{\beta }_{B } {d_R}^{B \dot\alpha }_{} {\phi}^{}_{}$ }, 
\textcolor{red}{ $\epsilon^{}_{\alpha \beta } \epsilon^{}_{\dot\alpha \dot\beta } {Q_L^\dagger}^{\dot\beta }_{B a } {Q_L}^{\alpha A a }_{} {d_R^\dagger}^{\beta }_{A } {d_R}^{B \dot\alpha }_{} {\phi}^{}_{}$ }, 
\textcolor{red}{ $\epsilon^{}_{\alpha \beta } \epsilon^{}_{\dot\alpha \dot\beta } {Q_L^\dagger}^{\dot\beta }_{A a } {Q_L^\dagger}^{\dot\alpha }_{B b } {Q_L}^{\alpha A a }_{} {Q_L}^{\beta B b }_{} {\phi}^{}_{}$ }, 
\textcolor{red}{ $\epsilon^{}_{\alpha \beta } \epsilon^{}_{\dot\alpha \dot\beta } {Q_L^\dagger}^{\dot\beta }_{B a } {Q_L^\dagger}^{\dot\alpha }_{A b } {Q_L}^{\alpha A a }_{} {Q_L}^{\beta B b }_{} {\phi}^{}_{}$ }, 
\textcolor{ForestGreen}{ $\epsilon^{}_{\dot\alpha \dot\beta } \epsilon^{}_{\dot\gamma \dot\delta } \epsilon^{a b }_{} {Q_L^\dagger}^{\dot\delta }_{A a } {Q_L^\dagger}^{\dot\gamma }_{B b } {u_R}^{A \dot\beta }_{} {d_R}^{B \dot\alpha }_{} {\phi}^{}_{}$ }, 
\textcolor{ForestGreen}{ $\epsilon^{}_{\dot\alpha \dot\beta } \epsilon^{}_{\dot\gamma \dot\delta } \epsilon^{a b }_{} {Q_L^\dagger}^{\dot\delta }_{B a } {Q_L^\dagger}^{\dot\beta }_{A b } {u_R}^{A \dot\alpha }_{} {d_R}^{B \dot\gamma }_{} {\phi}^{}_{}$ }, 
\textcolor{ForestGreen}{ $\epsilon^{}_{\alpha \beta } \epsilon^{}_{\gamma \delta } \epsilon^{}_{a b } {Q_L}^{\alpha A a }_{} {Q_L}^{\beta B b }_{} {u_R^\dagger}^{\delta }_{A } {d_R^\dagger}^{\gamma }_{B } {\phi}^{}_{}$ }, 
\textcolor{ForestGreen}{ $\epsilon^{}_{\alpha \beta } \epsilon^{}_{\gamma \delta } \epsilon^{}_{a b } {Q_L}^{\gamma A a }_{} {Q_L}^{\beta B b }_{} {u_R^\dagger}^{\alpha }_{B } {d_R^\dagger}^{\delta }_{A } {\phi}^{}_{}$ }, 
\textcolor{red}{ $\epsilon^{}_{\alpha \beta } \epsilon^{}_{\dot\alpha \dot\beta } {Q_L^\dagger}^{\dot\beta }_{A a } {Q_L}^{\alpha A a }_{} {u_R^\dagger}^{\beta }_{B } {u_R}^{B \dot\alpha }_{} {\phi}^{}_{}$ }, 
\textcolor{red}{ $\epsilon^{}_{\alpha \beta } \epsilon^{}_{\dot\alpha \dot\beta } {Q_L^\dagger}^{\dot\beta }_{B a } {Q_L}^{\alpha A a }_{} {u_R^\dagger}^{\beta }_{A } {u_R}^{B \dot\alpha }_{} {\phi}^{}_{}$ }, 
\textcolor{red}{ $\epsilon^{}_{\alpha \beta } \epsilon^{}_{\dot\alpha \dot\beta } {u_R^\dagger}^{\alpha }_{A } {u_R}^{A \dot\beta }_{} {d_R^\dagger}^{\beta }_{B } {d_R}^{B \dot\alpha }_{} {\phi}^{}_{}$ }, 
\textcolor{red}{ $\epsilon^{}_{\alpha \beta } \epsilon^{}_{\dot\alpha \dot\beta } {u_R^\dagger}^{\alpha }_{B } {u_R}^{A \dot\beta }_{} {d_R^\dagger}^{\beta }_{A } {d_R}^{B \dot\alpha }_{} {\phi}^{}_{}$ }, 
\textcolor{red}{ $\epsilon^{}_{\alpha \beta } \epsilon^{}_{\dot\alpha \dot\beta } {u_R^\dagger}^{\alpha }_{A } {u_R^\dagger}^{\beta }_{B } {u_R}^{A \dot\beta }_{} {u_R}^{B \dot\alpha }_{} {\phi}^{}_{}$ }, 
\textcolor{ForestGreen}{ $\epsilon^{\alpha \beta }_{} {B}^{}_{\gamma \beta } D_{\alpha \dot\alpha} {d_R^\dagger}^{\gamma }_{A } {d_R}^{A \dot\alpha }_{} {\phi}^{}_{}$ }, 
\textcolor{ForestGreen}{ $\epsilon^{\dot\alpha \dot\beta }_{} {B^\dagger}^{}_{\dot\beta \dot\gamma } {d_R^\dagger}^{\alpha }_{A } D_{\alpha \dot\alpha} {d_R}^{A \dot\gamma }_{} {\phi}^{}_{}$ }, 
\textcolor{ForestGreen}{ $\epsilon^{\alpha \beta }_{} {B}^{}_{\gamma \beta } D_{\alpha \dot\alpha} {e_R^\dagger}^{\gamma }_{} {e_R}^{\dot\alpha }_{} {\phi}^{}_{}$ }, 
\textcolor{ForestGreen}{ $\epsilon^{\dot\alpha \dot\beta }_{} {B^\dagger}^{}_{\dot\beta \dot\gamma } {e_R^\dagger}^{\alpha }_{} D_{\alpha \dot\alpha} {e_R}^{\dot\gamma }_{} {\phi}^{}_{}$ }, 
\textcolor{ForestGreen}{ $\epsilon^{\alpha \beta }_{} {B}^{}_{\gamma \beta } {L_L^\dagger}^{\dot\alpha }_{a } D_{\alpha \dot\alpha} {L_L}^{\gamma a }_{} {\phi}^{}_{}$ }, 
\textcolor{ForestGreen}{ $\epsilon^{\dot\alpha \dot\beta }_{} {B^\dagger}^{}_{\dot\beta \dot\gamma } D_{\alpha \dot\alpha} {L_L^\dagger}^{\dot\gamma }_{a } {L_L}^{\alpha a }_{} {\phi}^{}_{}$ }, 
\textcolor{ForestGreen}{ $\epsilon^{\alpha \beta }_{} {B}^{}_{\gamma \beta } {Q_L^\dagger}^{\dot\alpha }_{A a } D_{\alpha \dot\alpha} {Q_L}^{\gamma A a }_{} {\phi}^{}_{}$ }, 
\textcolor{ForestGreen}{ $\epsilon^{\dot\alpha \dot\beta }_{} {B^\dagger}^{}_{\dot\beta \dot\gamma } D_{\alpha \dot\alpha} {Q_L^\dagger}^{\dot\gamma }_{A a } {Q_L}^{\alpha A a }_{} {\phi}^{}_{}$ }, 
\textcolor{ForestGreen}{ $\epsilon^{\alpha \beta }_{} {B}^{}_{\gamma \beta } D_{\alpha \dot\alpha} {u_R^\dagger}^{\gamma }_{A } {u_R}^{A \dot\alpha }_{} {\phi}^{}_{}$ }, 
\textcolor{ForestGreen}{ $\epsilon^{\dot\alpha \dot\beta }_{} {B^\dagger}^{}_{\dot\beta \dot\gamma } {u_R^\dagger}^{\alpha }_{A } D_{\alpha \dot\alpha} {u_R}^{A \dot\gamma }_{} {\phi}^{}_{}$ }, 
\textcolor{BurntOrange}{ $ D_{\alpha \dot\alpha} {H^\dagger}^{}_{a } {H}^{a }_{} {d_R^\dagger}^{\alpha }_{A } {d_R}^{A \dot\alpha }_{} {\phi}^{}_{}$ }, 
\textcolor{BurntOrange}{ $ {H^\dagger}^{}_{a } D_{\alpha \dot\alpha} {H}^{a }_{} {d_R^\dagger}^{\alpha }_{A } {d_R}^{A \dot\alpha }_{} {\phi}^{}_{}$ }, 
\textcolor{BurntOrange}{ $ D_{\alpha \dot\alpha} {H^\dagger}^{}_{a } {H}^{a }_{} {e_R^\dagger}^{\alpha }_{} {e_R}^{\dot\alpha }_{} {\phi}^{}_{}$ }, 
\textcolor{BurntOrange}{ $ {H^\dagger}^{}_{a } D_{\alpha \dot\alpha} {H}^{a }_{} {e_R^\dagger}^{\alpha }_{} {e_R}^{\dot\alpha }_{} {\phi}^{}_{}$ }, 
\textcolor{BurntOrange}{ $ D_{\alpha \dot\alpha} {H^\dagger}^{}_{a } {H}^{a }_{} {L_L^\dagger}^{\dot\alpha }_{b } {L_L}^{\alpha b }_{} {\phi}^{}_{}$ }, 
\textcolor{BurntOrange}{ $ D_{\alpha \dot\alpha} {H^\dagger}^{}_{b } {H}^{a }_{} {L_L^\dagger}^{\dot\alpha }_{a } {L_L}^{\alpha b }_{} {\phi}^{}_{}$ }, 
\textcolor{BurntOrange}{ $ {H^\dagger}^{}_{a } D_{\alpha \dot\alpha} {H}^{a }_{} {L_L^\dagger}^{\dot\alpha }_{b } {L_L}^{\alpha b }_{} {\phi}^{}_{}$ }, 
\textcolor{BurntOrange}{ $ {H^\dagger}^{}_{b } D_{\alpha \dot\alpha} {H}^{a }_{} {L_L^\dagger}^{\dot\alpha }_{a } {L_L}^{\alpha b }_{} {\phi}^{}_{}$ }, 
\textcolor{BurntOrange}{ $ D_{\alpha \dot\alpha} {H^\dagger}^{}_{a } {H}^{a }_{} {Q_L^\dagger}^{\dot\alpha }_{A b } {Q_L}^{\alpha A b }_{} {\phi}^{}_{}$ }, 
\textcolor{BurntOrange}{ $ D_{\alpha \dot\alpha} {H^\dagger}^{}_{b } {H}^{a }_{} {Q_L^\dagger}^{\dot\alpha }_{A a } {Q_L}^{\alpha A b }_{} {\phi}^{}_{}$ }, 
\textcolor{BurntOrange}{ $ {H^\dagger}^{}_{a } D_{\alpha \dot\alpha} {H}^{a }_{} {Q_L^\dagger}^{\dot\alpha }_{A b } {Q_L}^{\alpha A b }_{} {\phi}^{}_{}$ }, 
\textcolor{BurntOrange}{ $ {H^\dagger}^{}_{b } D_{\alpha \dot\alpha} {H}^{a }_{} {Q_L^\dagger}^{\dot\alpha }_{A a } {Q_L}^{\alpha A b }_{} {\phi}^{}_{}$ }, 
\textcolor{BurntOrange}{ $\epsilon^{}_{a b } D_{\alpha \dot\alpha} {H}^{a }_{} {H}^{b }_{} {u_R^\dagger}^{\alpha }_{A } {d_R}^{A \dot\alpha }_{} {\phi}^{}_{}$ }, 
\textcolor{BurntOrange}{ $\epsilon^{a b }_{} D_{\alpha \dot\alpha} {H^\dagger}^{}_{a } {H^\dagger}^{}_{b } {u_R}^{A \dot\alpha }_{} {d_R^\dagger}^{\alpha }_{A } {\phi}^{}_{}$ }, 
\textcolor{BurntOrange}{ $ D_{\alpha \dot\alpha} {H^\dagger}^{}_{a } {H}^{a }_{} {u_R^\dagger}^{\alpha }_{A } {u_R}^{A \dot\alpha }_{} {\phi}^{}_{}$ }, 
\textcolor{BurntOrange}{ $ {H^\dagger}^{}_{a } D_{\alpha \dot\alpha} {H}^{a }_{} {u_R^\dagger}^{\alpha }_{A } {u_R}^{A \dot\alpha }_{} {\phi}^{}_{}$ }, 
\textcolor{ForestGreen}{ $\epsilon^{\alpha \beta }_{} D_{\beta \dot\alpha} {d_R^\dagger}^{\gamma }_{B } {d_R}^{A \dot\alpha }_{} {G}^{B }_{\gamma \alpha A } {\phi}^{}_{}$ }, 
\textcolor{ForestGreen}{ $\epsilon^{\dot\alpha \dot\beta }_{} {d_R^\dagger}^{\alpha }_{B } D_{\alpha \dot\beta} {d_R}^{A \dot\gamma }_{} {G^\dagger}^{B }_{\dot\alpha \dot\gamma A } {\phi}^{}_{}$ }, 
\textcolor{ForestGreen}{ $\epsilon^{\alpha \beta }_{} {L_L^\dagger}^{\dot\alpha }_{b } D_{\beta \dot\alpha} {L_L}^{\gamma a }_{} {W}^{b }_{\gamma \alpha a } {\phi}^{}_{}$ }, 
\textcolor{ForestGreen}{ $\epsilon^{\dot\alpha \dot\beta }_{} D_{\alpha \dot\beta} {L_L^\dagger}^{\dot\gamma }_{b } {L_L}^{\alpha a }_{} {W^\dagger}^{b }_{\dot\alpha \dot\gamma a } {\phi}^{}_{}$ }, 
\textcolor{ForestGreen}{ $\epsilon^{\alpha \beta }_{} {Q_L^\dagger}^{\dot\alpha }_{B a } D_{\beta \dot\alpha} {Q_L}^{\gamma A a }_{} {G}^{B }_{\gamma \alpha A } {\phi}^{}_{}$ }, 
\textcolor{ForestGreen}{ $\epsilon^{\dot\alpha \dot\beta }_{} D_{\alpha \dot\beta} {Q_L^\dagger}^{\dot\gamma }_{B a } {Q_L}^{\alpha A a }_{} {G^\dagger}^{B }_{\dot\alpha \dot\gamma A } {\phi}^{}_{}$ }, 
\textcolor{ForestGreen}{ $\epsilon^{\alpha \beta }_{} {Q_L^\dagger}^{\dot\alpha }_{A b } D_{\beta \dot\alpha} {Q_L}^{\gamma A a }_{} {W}^{b }_{\gamma \alpha a } {\phi}^{}_{}$ }, 
\textcolor{ForestGreen}{ $\epsilon^{\dot\alpha \dot\beta }_{} D_{\alpha \dot\beta} {Q_L^\dagger}^{\dot\gamma }_{A b } {Q_L}^{\alpha A a }_{} {W^\dagger}^{b }_{\dot\alpha \dot\gamma a } {\phi}^{}_{}$ }, 
\textcolor{ForestGreen}{ $\epsilon^{\alpha \beta }_{} D_{\beta \dot\alpha} {u_R^\dagger}^{\gamma }_{B } {u_R}^{A \dot\alpha }_{} {G}^{B }_{\gamma \alpha A } {\phi}^{}_{}$ }, 
\textcolor{ForestGreen}{ $\epsilon^{\dot\alpha \dot\beta }_{} {u_R^\dagger}^{\alpha }_{B } D_{\alpha \dot\beta} {u_R}^{A \dot\gamma }_{} {G^\dagger}^{B }_{\dot\alpha \dot\gamma A } {\phi}^{}_{}$ }, 
\textcolor{ForestGreen}{ $\epsilon^{\alpha \beta }_{} \epsilon^{\gamma \delta }_{} \epsilon^{\dot\alpha \dot\beta }_{} D_{\delta \dot\beta} {H^\dagger}^{}_{a } D_{\alpha \dot\alpha} {H}^{a }_{} {B}^{}_{\beta \gamma } {\phi}^{}_{}$ }, 
\textcolor{ForestGreen}{ $\epsilon^{\alpha \beta }_{} \epsilon^{\dot\alpha \dot\beta }_{} \epsilon^{\dot\gamma \dot\delta }_{} D_{\alpha \dot\delta} {H^\dagger}^{}_{a } D_{\beta \dot\beta} {H}^{a }_{} {B^\dagger}^{}_{\dot\alpha \dot\gamma } {\phi}^{}_{}$ }, 
\textcolor{ForestGreen}{ $\epsilon^{\alpha \beta }_{} \epsilon^{\dot\alpha \dot\beta }_{} {H^\dagger}^{}_{a } {H^\dagger}^{}_{b } D_{\alpha \dot\beta} {H}^{a }_{} D_{\beta \dot\alpha} {H}^{b }_{} {\phi}^{}_{}$ }, 
\textcolor{ForestGreen}{ $\epsilon^{\alpha \beta }_{} \epsilon^{\dot\alpha \dot\beta }_{} D_{\alpha \dot\beta} {H^\dagger}^{}_{b } D_{\beta \dot\alpha} {H^\dagger}^{}_{a } {H}^{a }_{} {H}^{b }_{} {\phi}^{}_{}$ }, 
\textcolor{red}{ $\epsilon^{\alpha \beta }_{} \epsilon^{\dot\alpha \dot\beta }_{} D_{\alpha \dot\beta} {H^\dagger}^{}_{a } {H^\dagger}^{}_{b } D_{\beta \dot\alpha} {H}^{a }_{} {H}^{b }_{} {\phi}^{}_{}$ }, 
\textcolor{red}{ $\epsilon^{\alpha \beta }_{} \epsilon^{\dot\alpha \dot\beta }_{} D_{\alpha \dot\beta} {H^\dagger}^{}_{a } D_{\beta \dot\alpha} {H}^{a }_{} {\phi}^{}_{} {\phi}^{}_{} {\phi}^{}_{}$ }, 
\textcolor{ForestGreen}{ $\epsilon^{\alpha \beta }_{} \epsilon^{\gamma \delta }_{} \epsilon^{\dot\alpha \dot\beta }_{} D_{\delta \dot\beta} {H^\dagger}^{}_{b } D_{\alpha \dot\alpha} {H}^{a }_{} {W}^{b }_{\beta \gamma a } {\phi}^{}_{}$ }, 
\textcolor{ForestGreen}{ $\epsilon^{\alpha \beta }_{} \epsilon^{\dot\alpha \dot\beta }_{} \epsilon^{\dot\gamma \dot\delta }_{} D_{\alpha \dot\delta} {H^\dagger}^{}_{b } D_{\beta \dot\beta} {H}^{a }_{} {W^\dagger}^{b }_{\dot\alpha \dot\gamma a } {\phi}^{}_{}$ }, 
\textcolor{BurntOrange}{ $\epsilon^{\dot\alpha \dot\beta }_{} {H^\dagger}^{}_{a } D_{\beta \dot\beta} {e_R^\dagger}^{\alpha }_{} D_{\alpha \dot\alpha} {L_L}^{\beta a }_{} {\phi}^{}_{}$ }, 
\textcolor{BurntOrange}{ $\epsilon^{\dot\alpha \dot\beta }_{} {H^\dagger}^{}_{a } D_{\beta \dot\beta} {e_R^\dagger}^{\alpha }_{} D_{\alpha \dot\alpha} {L_L}^{\beta a }_{} {\phi}^{}_{}$ }, 
\textcolor{BurntOrange}{ $\epsilon^{\alpha \beta }_{} {H}^{a }_{} D_{\alpha \dot\beta} {e_R}^{\dot\alpha }_{} D_{\beta \dot\alpha} {L_L^\dagger}^{\dot\beta }_{a } {\phi}^{}_{}$ }, 
\textcolor{BurntOrange}{ $\epsilon^{\alpha \beta }_{} {H}^{a }_{} D_{\alpha \dot\beta} {e_R}^{\dot\alpha }_{} D_{\beta \dot\alpha} {L_L^\dagger}^{\dot\beta }_{a } {\phi}^{}_{}$ }, 
\textcolor{BurntOrange}{ $\epsilon^{\alpha \beta }_{} {H}^{a }_{} D_{\alpha \dot\beta} {Q_L^\dagger}^{\dot\alpha }_{A a } D_{\beta \dot\alpha} {d_R}^{A \dot\beta }_{} {\phi}^{}_{}$ }, 
\textcolor{BurntOrange}{ $\epsilon^{\alpha \beta }_{} {H}^{a }_{} D_{\alpha \dot\beta} {Q_L^\dagger}^{\dot\alpha }_{A a } D_{\beta \dot\alpha} {d_R}^{A \dot\beta }_{} {\phi}^{}_{}$ }, 
\textcolor{BurntOrange}{ $\epsilon^{\dot\alpha \dot\beta }_{} {H^\dagger}^{}_{a } D_{\beta \dot\beta} {Q_L}^{\alpha A a }_{} D_{\alpha \dot\alpha} {d_R^\dagger}^{\beta }_{A } {\phi}^{}_{}$ }, 
\textcolor{BurntOrange}{ $\epsilon^{\dot\alpha \dot\beta }_{} {H^\dagger}^{}_{a } D_{\beta \dot\beta} {Q_L}^{\alpha A a }_{} D_{\alpha \dot\alpha} {d_R^\dagger}^{\beta }_{A } {\phi}^{}_{}$ }, 
\textcolor{BurntOrange}{ $\epsilon^{\alpha \beta }_{} \epsilon^{a b }_{} {H^\dagger}^{}_{a } D_{\alpha \dot\beta} {Q_L^\dagger}^{\dot\alpha }_{A b } D_{\beta \dot\alpha} {u_R}^{A \dot\beta }_{} {\phi}^{}_{}$ }, 
\textcolor{BurntOrange}{ $\epsilon^{\alpha \beta }_{} \epsilon^{a b }_{} {H^\dagger}^{}_{a } D_{\alpha \dot\beta} {Q_L^\dagger}^{\dot\alpha }_{A b } D_{\beta \dot\alpha} {u_R}^{A \dot\beta }_{} {\phi}^{}_{}$ }, 
\textcolor{BurntOrange}{ $\epsilon^{\dot\alpha \dot\beta }_{} \epsilon^{}_{a b } {H}^{a }_{} D_{\beta \dot\beta} {Q_L}^{\alpha A b }_{} D_{\alpha \dot\alpha} {u_R^\dagger}^{\beta }_{A } {\phi}^{}_{}$ }, 
\textcolor{BurntOrange}{ $\epsilon^{\dot\alpha \dot\beta }_{} \epsilon^{}_{a b } {H}^{a }_{} D_{\beta \dot\beta} {Q_L}^{\alpha A b }_{} D_{\alpha \dot\alpha} {u_R^\dagger}^{\beta }_{A } {\phi}^{}_{}$ }

\line(1,0){40}
 
\textcolor{blue}{ $\epsilon^{}_{\alpha \beta } \epsilon^{}_{a b } \epsilon^{}_{d e } {H}^{d }_{} {H}^{b }_{} {L_L}^{\alpha a }_{} {L_L}^{\beta e }_{} {\phi}^{}_{} {\phi}^{}_{}$ }, 
\textcolor{blue}{ $\epsilon^{}_{\dot\alpha \dot\beta } \epsilon^{a b }_{} \epsilon^{d e }_{} {H^\dagger}^{}_{d } {H^\dagger}^{}_{b } {L_L^\dagger}^{\dot\beta }_{a } {L_L^\dagger}^{\dot\alpha }_{e } {\phi}^{}_{} {\phi}^{}_{}$ }, 
\textcolor{ForestGreen}{ $\epsilon^{}_{\alpha \beta } \epsilon^{}_{\dot\alpha \dot\beta } \epsilon^{}_{a b } \epsilon^{}_{A B C } {e_R}^{\dot\beta }_{} {Q_L}^{\alpha B a }_{} {Q_L}^{\beta C b }_{} {u_R}^{A \dot\alpha }_{} {\phi}^{}_{}$ }, 
\textcolor{ForestGreen}{ $\epsilon^{}_{\alpha \beta } \epsilon^{}_{\dot\alpha \dot\beta } \epsilon^{a b }_{} \epsilon^{A B C }_{} {e_R^\dagger}^{\alpha }_{} {Q_L^\dagger}^{\dot\beta }_{A a } {Q_L^\dagger}^{\dot\alpha }_{B b } {u_R^\dagger}^{\beta }_{C } {\phi}^{}_{}$ }, 
\textcolor{ForestGreen}{ $\epsilon^{}_{\dot\alpha \dot\beta } \epsilon^{}_{\dot\gamma \dot\delta } \epsilon^{}_{A B C } {e_R}^{\dot\delta }_{} {u_R}^{B \dot\beta }_{} {u_R}^{C \dot\gamma }_{} {d_R}^{A \dot\alpha }_{} {\phi}^{}_{}$ }, 
\textcolor{ForestGreen}{ $\epsilon^{}_{\alpha \beta } \epsilon^{}_{\gamma \delta } \epsilon^{A B C }_{} {e_R^\dagger}^{\alpha }_{} {u_R^\dagger}^{\delta }_{A } {u_R^\dagger}^{\beta }_{B } {d_R^\dagger}^{\gamma }_{C } {\phi}^{}_{}$ }, 
\textcolor{ForestGreen}{ $\epsilon^{}_{\alpha \beta } \epsilon^{}_{\gamma \delta } \epsilon^{}_{a b } \epsilon^{}_{d e } \epsilon^{}_{A B C } {L_L}^{\gamma a }_{} {Q_L}^{\beta B e }_{} {Q_L}^{\alpha C b }_{} {Q_L}^{\delta A d }_{} {\phi}^{}_{}$ }, 
\textcolor{ForestGreen}{ $\epsilon^{}_{\dot\alpha \dot\beta } \epsilon^{}_{\dot\gamma \dot\delta } \epsilon^{a b }_{} \epsilon^{d e }_{} \epsilon^{A B C }_{} {L_L^\dagger}^{\dot\delta }_{a } {Q_L^\dagger}^{\dot\beta }_{A e } {Q_L^\dagger}^{\dot\alpha }_{B b } {Q_L^\dagger}^{\dot\gamma }_{C d } {\phi}^{}_{}$ }, 
\textcolor{ForestGreen}{ $\epsilon^{}_{\alpha \beta } \epsilon^{}_{\dot\alpha \dot\beta } \epsilon^{}_{a b } \epsilon^{}_{A B C } {L_L}^{\alpha a }_{} {Q_L}^{\beta B b }_{} {u_R}^{C \dot\beta }_{} {d_R}^{A \dot\alpha }_{} {\phi}^{}_{}$ }, 
\textcolor{ForestGreen}{ $\epsilon^{}_{\alpha \beta } \epsilon^{}_{\dot\alpha \dot\beta } \epsilon^{a b }_{} \epsilon^{A B C }_{} {L_L^\dagger}^{\dot\beta }_{a } {Q_L^\dagger}^{\dot\alpha }_{A b } {u_R^\dagger}^{\alpha }_{B } {d_R^\dagger}^{\beta }_{C } {\phi}^{}_{}$ }
\end{description}
\end{adjustwidth*}
Upper case Latin indices transform under \suc, lower case Latin under \sul, undotted Greek under the left-handed part of the Lorentz group, and dotted Greek indices under the right-handed part. Horizontal lines separate operators that conserve both $B$ and $L$, and those that don't. Note that the inclusion of a real scalar singlet spoils the correspondence between operator dimension and $B$ and $L$ charges which is present in the Standard Model \cite{deGouvea:2014lva,Kobach:2016ami}.

We use colours to denote the effects of requiring that the lagrangian respect CP, in the two cases that $\phi$ is CP even and CP odd. Pairs of adjacent operators in \textcolor{blue}{blue} are Hermitian conjugates of each other (up to a sign): requiring CP invariance, these operators must have equal real coefficients (up to the same sign). Pairs of adjacent operators in \textcolor{ForestGreen}{green} or \textcolor{BurntOrange}{orange} are also Hermitian conjugates of each other (up to a sign): requiring CP invariance, these operators must have either equal real or equal imaginary coefficients (up to the same sign) in the respective cases of $\phi$ CP even and CP odd when \textcolor{ForestGreen}{green}, or in the respective cases of $\phi$ CP odd and CP even when \textcolor{BurntOrange}{orange}. \textcolor{red}{Red} operators are simply forbidden if we require CP invariance and that $\phi$ be CP odd.


\section{Comparison with previous results \label{sec:compare}}

At dimension 5, Ref. \cite{Franceschini:2016gxv} has two extra operators in their basis. In our notation, these may be written as $\epsilon^{\al \be} \epsilon^{\dot\al \dot\be} \phi D_{\al \dot\be} \phi D_{\be \dot \al} \phi$ and $\epsilon^{\al \be} \epsilon^{\dot\al \dot\be} \phi D_{\al \dot\be} {H^\dagger}_a D_{\be \dot \al} H^a$. To see that the first is redundant, we first integrate by parts to get a `$\phi^2 \Box \phi$' term and then use the equation of motion for $\phi$ to obtain a sum of products of $\phi$ and $H$ fields, without derivatives. To see that the second is redundant, first integrate by parts to obtain a sum of three terms: `$\phi H^\dagger D^2 H$', `$\Box \phi H^\dagger H$', and `$\pt_\mu \phi H^\dagger \overleftrightarrow{D}^\mu H $'. The former two operators may be expressed in terms of others in the basis using the equations of motion for the Higgs and singlet respectively. The latter term, via the $B$'s equation of motion, may be expressed as a sum of `$\pt_\mu \phi \bar{\Psi} \gamma^\mu \Psi$'-like terms, where $\Psi$ is a generic fermion. Finally, integrating by parts again, and using the fermions' equations of motion, we obtain a sum of terms already in the basis.

Note that, at dimension 6, Ref. \cite{Franceschini:2016gxv} opt to exclude the lepton-number violating operator obtained by multiplying the Weinberg operator \cite{PhysRevLett.43.1566} in the SM at dimension 5 by $\phi$, viz. $\epsilon^{}_{\alpha \beta } \epsilon^{}_{a b } \epsilon^{}_{d e } {H}^{d }_{} {H}^{b }_{} {L_L}^{\alpha a }_{} {L_L}^{\beta e }_{} {\phi}^{}_{}$ and its Hermitian conjugate.

\section*{Acknowledgements}
We thank T.~Melia, M.~Bauer and M.~Neubert for pointing out errors in an earlier version. BG acknowledges the support of STFC (grant ST/L000385/1) and King's College, Cambridge. DS also acknowledges the support of STFC, and Emmanuel College, Cambridge, and thanks KITP for hospitality. This research was supported in part by the National Science Foundation under Grant No. PHY11-25915.

\bibliography{operators.bib}

\providecommand{\href}[2]{#2}\begingroup\raggedright\begin{thebibliography}{10}

\bibitem{atlasDiphoton}
T.~A. collaboration, \emph{{Search for resonances decaying to photon pairs in
  3.2 fb$^{-1}$ of $pp$ collisions at $\sqrt{s}$ = 13 TeV with the ATLAS
  detector}}, .

\bibitem{CMS:2015dxe}
{\scshape CMS} collaboration, C.~Collaboration, \emph{{Search for new physics
  in high mass diphoton events in proton-proton collisions at 13TeV}}, .

\bibitem{Franceschini:2016gxv}
R.~Franceschini, G.~F. Giudice, J.~F. Kamenik, M.~McCullough, F.~Riva,
  A.~Strumia et~al., \emph{{Digamma, what next?}},
  \href{http://arxiv.org/abs/1604.06446}{{\tt 1604.06446}}.

\bibitem{Lehman:2015via}
L.~Lehman and A.~Martin, \emph{{Hilbert Series for Constructing Lagrangians:
  expanding the phenomenologist's toolbox}},
  \href{http://dx.doi.org/10.1103/PhysRevD.91.105014}{\emph{Phys. Rev.} {\bf
  D91} (2015) 105014}, [\href{http://arxiv.org/abs/1503.07537}{{\tt
  1503.07537}}].

\bibitem{Henning:2015daa}
B.~Henning, X.~Lu, T.~Melia and H.~Murayama, \emph{{Hilbert series and operator
  bases with derivatives in effective field theories}},
  \href{http://arxiv.org/abs/1507.07240}{{\tt 1507.07240}}.

\bibitem{Lehman:2015coa}
L.~Lehman and A.~Martin, \emph{{Low-derivative operators of the Standard Model
  effective field theory via Hilbert series methods}},
  \href{http://dx.doi.org/10.1007/JHEP02(2016)081}{\emph{JHEP} {\bf 02} (2016)
  081}, [\href{http://arxiv.org/abs/1510.00372}{{\tt 1510.00372}}].

\bibitem{Henning:2015alf}
B.~Henning, X.~Lu, T.~Melia and H.~Murayama, \emph{{2, 84, 30, 993, 560, 15456,
  11962, 261485, ...: Higher dimension operators in the SM EFT}},
  \href{http://arxiv.org/abs/1512.03433}{{\tt 1512.03433}}.

\bibitem{deft}
B.~Gripaios and D.~Sutherland, \emph{to appear}, .

\bibitem{Dreiner:2008tw}
H.~K. Dreiner, H.~E. Haber and S.~P. Martin, \emph{{Two-component spinor
  techniques and Feynman rules for quantum field theory and supersymmetry}},
  \href{http://dx.doi.org/10.1016/j.physrep.2010.05.002}{\emph{Phys. Rept.}
  {\bf 494} (2010) 1--196}, [\href{http://arxiv.org/abs/0812.1594}{{\tt
  0812.1594}}].

\bibitem{Grzadkowski:2010es}
B.~Grzadkowski, M.~Iskrzynski, M.~Misiak and J.~Rosiek, \emph{{Dimension-Six
  Terms in the Standard Model Lagrangian}},
  \href{http://dx.doi.org/10.1007/JHEP10(2010)085}{\emph{JHEP} {\bf 10} (2010)
  085}, [\href{http://arxiv.org/abs/1008.4884}{{\tt 1008.4884}}].

\bibitem{Einhorn:2013kja}
M.~B. Einhorn and J.~Wudka, \emph{{The Bases of Effective Field Theories}},
  \href{http://dx.doi.org/10.1016/j.nuclphysb.2013.08.023}{\emph{Nucl. Phys.}
  {\bf B876} (2013) 556--574}, [\href{http://arxiv.org/abs/1307.0478}{{\tt
  1307.0478}}].

\bibitem{PhysRevLett.43.1566}
S.~Weinberg, \emph{Baryon- and lepton-nonconserving processes},
  \href{http://dx.doi.org/10.1103/PhysRevLett.43.1566}{\emph{Phys. Rev. Lett.}
  {\bf 43} (Nov, 1979) 1566--1570}.

\bibitem{deGouvea:2014lva}
A.~de~Gouvea, J.~Herrero-Garcia and A.~Kobach, \emph{{Neutrino Masses, Grand
  Unification, and Baryon Number Violation}},
  \href{http://dx.doi.org/10.1103/PhysRevD.90.016011}{\emph{Phys. Rev.} {\bf
  D90} (2014) 016011}, [\href{http://arxiv.org/abs/1404.4057}{{\tt
  1404.4057}}].

\bibitem{Kobach:2016ami}
A.~Kobach, \emph{{Baryon Number, Lepton Number, and Operator Dimension in the
  Standard Model}},
  \href{http://dx.doi.org/10.1016/j.physletb.2016.05.050}{\emph{Phys. Lett.}
  {\bf B758} (2016) 455--457}, [\href{http://arxiv.org/abs/1604.05726}{{\tt
  1604.05726}}].

\end{thebibliography}\endgroup
\bibliographystyle{JHEP}

\end{document}